# Epidemiologically optimal static networks from temporal network data

## Petter Holme

*Department of Energy Science, Sungkyunkwan University, 440-746 Suwon, Korea*

*IceLab, Department of Physics, Umeå University, 90187 Umeå, Sweden*

*Department of Sociology, Stockholm University, 10961 Stockholm, Sweden*

Network epidemiology's most important assumption is that the contact structure over which infectious diseases propagate can be represented as a static network. However, contacts are highly dynamic, changing at many time scales. In this paper, we investigate conceptually simple methods to construct static graphs for network epidemiology from temporal contact data. We evaluate these methods on empirical and synthetic model data. For almost all our cases, the network representation that captures most relevant information is a so-called exponential-threshold network. In these, each contact contributes with a weight decreasing exponentially with time, and there is an edge between a pair of vertices if the weight between them exceeds a threshold. Networks of aggregated contacts over an optimally chosen time window perform almost as good as the exponential-threshold networks. On the other hand, networks of accumulated contacts over the entire sampling time, and networks of concurrent partnerships, perform worse. We discuss these observations in the context of the temporal and topological structure of the data sets.

## Introduction

In the 1980's and 90's, mathematical epidemiology of infectious diseases made great progress. During these years, researchers went from models where every individual meets everyone else with equal probability, to a framework of networks where people are considered as connected if one can infect the other. This new body of theories, network epidemiology [1–4], has altered our understanding of disease spreading profoundly. For example, it has changed the concept of epidemic thresholds, outbreak diversity and the role of social networks in intervening infectious disease outbreaks. Furthermore, research in network epidemiology has produced many new techniques to analyze contact data [3–5], model disease spreading [3,4,6,7], discovering influential spreaders [8–10], detecting outbreaks [11], etc. Still, network epidemiology rests on coarse simplifications, perhaps the biggest being that that one usually does not explicitly model the dynamic aspects of contact patterns. If we consider two individuals, and assume one of them is infective (infected by a pathogen and able to spread it further), the probability of contagion between the two is in practice not constant in time. The changes in the chance of contagion happens at different time scales—from the order of decades, as people are born and die, to the order of minutes, as they come in and out of range for pathogen transmission. The situation becomes even more complicated if we consider an emerging disease outbreak. First, if we want to apply network epidemiology to control the spreading, we have to predict the future contacts, not just map out the past [10]. Second, the mechanisms behind how and when people make contacts may change from the fact that the people are aware of the epidemics [12,13]. However, the theory and methods to handle full contact patterns—including both temporal and topological information—is not at all as developed as static network epidemiology [4,14,15]. Therefore, static network epidemiology is still an important theoretical framework. To be able to study contact patterns by static methods, one needs to eliminate the temporal dimension. This can be done in many ways, and this paper aims at finding the optimal way.

Consider a sequence of contacts—triples $(i,j,t)$ carrying the information about when (assuming a discrete time $t$) pairs of individuals ($i$ and $j$) have been in contact. A good static network representation of such a contact sequence should give the same predictions about the disease dynamics as the contact sequence itself. The predictions we focus on in this paper are related to how influential an individual is in the disease spreading. Assuming a disease is introduced in a population by individual $i$, we compare the expected outbreak size $\Sigma_i$ in a simulation on the contact sequence with two static-network predictors of importance: $i$'s degree $k_i$ and coreness $c_i$ [8] (roughly speaking, the size of the most close-knit community around $i$). The better the performance of the static network predictors, the better is the network representation. The reason we focus on predicting the importance of individuals rather than e.g. the epidemic threshold is that it is less dependent on the parameters of the disease-spreading model. If, for a given set of parameter values, one can predict the total outbreak size but not the important disease spreaders, that is more likely a coincidence than if one can predict the important disease spreaders but not the outbreak size. This is important since, as our main focus is to scan different types of network representations, we will have to limit ourselves to a few parameter values of the disease spreading simulations.

In this paper, we will use both empirical and artificially generated temporal-network data sets. We investigate three classes of network representations to find which one that can predict $\Sigma_i$ from $k_i$ or $c_i$ the best. Furthermore, we investigate how the performance depends on the temporal and topological structures of the data.

# RESULTS

**Preliminaries**

We will compare three conceptually simple methods of reducing a contact sequence to a static network (illustrated in Fig. 1). In the first method, time-slice networks [16,17], an edge means that two vertices have been in contact within an interval $[t_{start}, t_{stop}]$. The second representation, concurrency networks [18], adds edges between pairs of vertices with contacts both before and after the interval $[t_{start}, t_{stop}]$. This is thus a network of edges, or relationships that are concurrently active over the time window. This method takes its name from literature of sexually transmitted infections where it is believed that the level of concurrent partnerships is a key-factor to understand how contact patterns influence epidemics [19,20]. The third method is exponential-threshold networks. In these, every contact contributes with a weight—decaying exponentially $e^{-t/\tau}$ with the time $t$ of the contact (so that later contacts are weighted less than earlier)—to the pairs of vertices. Then this network of weights is reduced to a simple graph by including edges for all weighted edges above a certain threshold $\Omega$. All three methods have two control parameter each—the first two methods having the endpoints of an interval as parameters, the last one has a decay parameter $\tau$ and a threshold $\Omega$. Note that other common network representations are limit cases of one of these methods. E.g., a network of accumulated contacts is the same as a time-slice network over the entire sampling time. A simple threshold network (where an edge represent all pairs of vertices with at least $\Omega$ contacts) is the $\tau \to \infty$ limit of the exponential-threshold networks. A more detailed description of the network representations can be found in the Methods section.

As mentioned, we evaluate the network representations by comparing the importance ($\Sigma_i$) of individuals for the disease spreading in a temporal network and the assessed importance ($k_i$ or $c_i$) from the derived static network. $\Sigma_i$ comes from a Susceptible–Infective–Removed (SIR) simulation (for details, see the Methods section). The stronger the correlation between $\Sigma_i$, on one hand, and $k_i$ or $c_i$ on the other, the better is the network representation. As it turns out $\Sigma_i$ and $k_i$, or $c_i$, typically have strongly nonlinear relationships, which makes the Pearson correlation coefficient less related to the actual predictability of the data. For this reason, and that Kendall's tau is prohibitively slow to compute in our case, we focus on the Spearman rank correlation. (We test the Pearson and Kendall coefficient for some of the data and find that all three coefficients rank the methods in the same order and are optimized for about the same parameter values.)

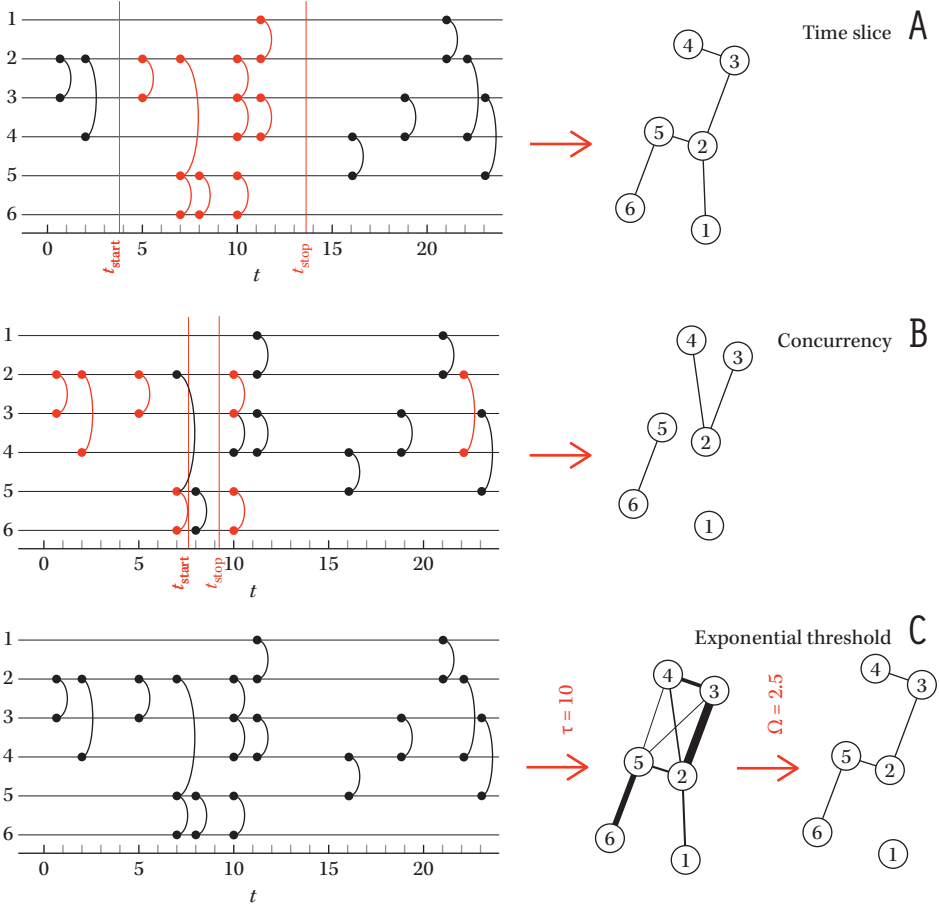

**Figure 1.** Illustration of the network representations. To the left in all panels is a temporal network where each horizontal line is the timeline of a vertex. The vertical curves symbolize the contact between two vertices as one timestep. Panel A shows the construction of the time-slice network. Two vertices are connected if they have at least one contact in the time interval $[t_{start}, t_{stop}]$. In panel B, a vertex pair is connected if they have contacts before $t_{start}$ and after $t_{stop}$. Finally, panel C illustrates how the contact sequence is reduced to a weighted graph that is converted to an unweighted graph by requiring an edge to have a weight over a certain threshold $\Omega$. The thickness of the lines in panel C is proportional to the weight between the pair.

**Table 1.** Empirical data sets—sizes and basic temporal statistics. $N$, $M$ and $L$ are the number of individuals, edges in the accumulated network (pairs with at least one contact) and contacts, respectively. $\lambda$ is the per-contact transmission probability used in the disease-spreading simulations. $T$ is the total sampling time. $B$ is the burstiness index of the entire set of interevent times between pairs of at least two contacts. The values for the *Gallery* data are averaged over all 69 days. The values in parentheses show the standard errors of the number in the order of its last decimal. For details of the definitions of parameters, see the Methods section.

|   | E-mail 1 | E-mail 2 | Dating | Gallery | Conference | Prostitution |
|---|---|---|---|---|---|---|
| $N$ | 57,189 | 3,188 | 28,972 | 159(8) | 113 | 16,730 |
| $M$ | 92,442 | 31,857 | 115,684 | 647(57) | 2,196 | 39,044 |
| $L$ | 444,160 | 115,684 | 529,890 | 6,027(350) | 20,818 | 50,632 |
| $\lambda$ | 0.298 | 0.031 | 0.108 | 0.067 | 0.028 | 0.416 |
| $T$ | 112.0d | 81.6d | 512.0d | 7.3(1)h | 2.5d | 2,232d |
| $B$ | 0.416 | 0.383 | 0.652 | 0.40(6) | 0.632 | 0.432 |

**Empirical networks**

As a start, we will analyze empirical contact sequences of the type outlined above (lists of potentially contagious contacts—who has been in contact with whom at what time). These empirical data sets are more or less related to disease spreading; but they all serve as examples of different temporal-network structures. The data sets fall into three categories—online communication, face-to-face and sexual encounters. The latter two categories are of course more interesting for the spread of infectious diseases (while the former perhaps could be interesting for the spread of e-mail viruses). Of online communication data, we study two e-mail networks—from Refs. [21] (*E-mail 1*) and [22] (*E-mail 2*). In these data sets, a sent e-mail represents a contact. Even though an e-mail is naturally directed, to analyze all the data in the same way, we treat it as undirected. The two e-mail data sets are sampled from a group of e-mail accounts. One difference between them is that the data of Ref. [21] includes contacts to external e-mail accounts while the data of Ref. [22] only records e-mails between the sampled accounts. One method is probably not better than the other. To avoid these boundary effects, one can study communication within a closed community. We do this with data from an Internet dating community first presented in Ref. [23] (*Dating*). The face-to-face data sets are gathered by radio-frequency identification sensors worn by the participants of a conference [24] (*Conference*) and visitors of a gallery [25] (*Gallery*). In these data sets, a contact is recorded, at 20 seconds intervals, if two individuals are within range (1–1.5m). Finally, we use a data set of sexual encounters gathered from a web forum where sex-buyers evaluate escorts [26] (*Prostitution*). We list some basic statistics of the data sets in Table 1.

Turning to the main results of this section, we display the performance of the network representations in Tables 2 (with degree as the importance measure in the static network) and 3 (with coreness as importance measure). The most discernable result is that the exponential-threshold networks have the highest score for all data sets and importance measures except one case (with degree as importance measure for the *Conference* data). Indeed, the Spearman $\rho$-values are all relatively high, meaning that important spreaders are highly predictable from just the contact patterns (although not possible to forecast, as this is a *post hoc* analysis). This suggests that the exponential-threshold representation is a good general way of constructing networks for network epidemiology (which we will argue for more below). Another observation is that the aggregate networks, the most common static network representation of temporal network data, perform very poorly (ranging from 51%–91% of the maximal corre-lation value). The concurrency networks perform very differently for different data sets—sometimes (*E-mail 2*) they are close to the best, sometimes (*Prostitution*, Table 3) remarkably bad. We note that the concurrency network representation is typically optimized for $t_{start} = t_{start}$ (the special case studied in Ref. [18]). I.e., longer concurrent partnerships (the set of contacts between a pair of vertices) does not predict disease spreaders better that the mere fact that they are concurrent. The occasional poor performance of the concurrency networks is a bit surprising in the light of the reported importance of concurrent partnerships for disease spreading in sexual networks [19,20]. An explanation could be that these studies concern population averages rather than the relative importance of individuals. The time-slice networks are performing consistently well—in one case better, and in the other cases close to the exponential-threshold networks (on average $\rho \approx 0.09$ lower). They have most relevant information if the time interval begins early. Indeed, the optimizing starting time is almost always the same as the beginning of the epidemics. This means they are also in practice, like the exponential-threshold networks, weighing the interactions with a weight decreasing with time (only that this weight function is discontinuous). The relative duration of the optimal time slice varies considerably (from 10% to 77% of the entire sampling time). Ref. [17] points out that time-slice networks of phone communications are most complex for intermediate time windows; perhaps our optimal time-slice networks coincide with this region. Comparing Tables 2 and 3, we see that the results are rather similar for the degree and coreness values. In most cases, coreness outperforms degree (confirming the conclusions of Ref. [8]), but the difference is often in the third decimal of $\rho_{max}$. We note that the optimal performance varies quite a bit—from 0.74 for the *Prostitution* data to 0.93 for *E-mail 2*.

We now take a deeper look at the regions of optimal parameter values for the three classes of network representations. If one wants a quick analysis without the optimization procedure of this paper, then how can one set the parameters? Are there rules of thumb? We use the *Prostitution* data as an example in Fig. 2. The other data sets behave qualitatively similar (with one exception, mentioned below). The window of the optimal time-slice networks starts, with few exceptions, at the same time as the first contact. In other words, the initial contacts of the seed and its surroundings are so important that the other early contacts (between vertices that are out of reach of the infection at that stage, and thus with a potentially negative contribution to the correlation coefficient) do not matter. However, the end of the interval is harder to guess. Presumably, this value should be of the order of the peak of the outbreak. After the peak, the contacts should have less influence on the disease evolution and thus on $\Sigma_i$. Since simulated disease should spread fast in dense data sets like *E-mail 2* and *Conference* (with about 36 and 184 contacts per individual on average, respectively), it is natural that these data sets show low tstop-values (relative both to the sampling time and the mean interevent time). Nevertheless, we still do not know how to estimate this value without running disease simulations. The good news is that the network representation is rather insensitive to the choice of tstop. The concurrent networks typically are maximized at $t_{start} \approx t_{stop}$ for some intermediate value smaller than the duration. Also here, it is hard to give an estimate of this parameter value, more than it happens within the optimal time window of the time-slice data. The last method, the exponential-threshold networks, is frequently optimized along a curve $\tau \sim e^{\Omega/\Omega'}$, where $\Omega'$ is a constant (and the functional form follows from Eq. (1)). This is because larger decay factors give larger weights and thus larger thresholds. The *Conference* data, however, is optimized for values close to the lower limit of decay exponent (which is linearly increasing with the threshold value). Our conclusion is that no matter which one of our three representations one use, one typically needs to optimize one, but not two, of its parameters—$t_{stop}$ for time-slice networks, $t_{start} = t_{stop}$ for concurrency networks and $\Omega'$ for the exponential-threshold networks.

**Table 2.** Maximal performance values for the empirical data sets using degree as importance measure for the static networks. The last column shows values for the network of accumulated contacts. For the *Gallery* data, the values are averaged over the 69 days. The values in parentheses represent the standard error in the order of the last digit. The largest values for each data set, for degree and coreness respectively, are highlighted with boldface. The parameters of temporal dimensions—$t_{start}$, $t_{stop}$ and $\tau$—are measured in units of the total sampling time $T$ of the respective data set (see Table 1).

| | | E-mail 1 | E-mail 2 | Dating | Gallery | Conference | Prostitution |
|---|---|---|---|---|---|---|---|
| Time slice | $\rho_{max}$ | 0.735(5) | 0.907(4) | 0.821(3) | 0.77(2) | **0.787(2)** | 0.711(2) |
| | $t_{start}$ | 0.0 | 0.0 | 0.0 | 0.0 | 0.0 | 0.0 |
| | $t_{stop}$ | 0.42(3) | 0.25(2) | 0.65(3) | 0.72(5) | 0.10(1) | 0.77(2) |
| Concurrency | $\rho_{max}$ | 0.497(5) | 0.914(1) | 0.421(3) | 0.53(2) | 0.743(2) | 0.301(4) |
| | $t_{start}$ | 0.25(3) | 0.20(3) | 0.25(2) | 0.39(2) | 0.10(3) | 0.60(2) |
| | $t_{stop}$ | 0.25(3) | 0.20(3) | 0.25(2) | 0.39(2) | 0.11(2) | 0.60(2) |
| Exponential threshold | $\rho_{max}$ | **0.771(2)** | **0.931(3)** | **0.861(2)** | **0.87(1)** | 0.775(2) | **0.721(3)** |
| | $\tau$ | 0.40(1) | 1.0(1) | 0.10(4) | 0.70(3) | 0.04(1) | 0.040(3) |
| | $\Omega$ | 0.30(2) | 0.26(2) | 0.16(2) | 0.71(2) | 0.020(2) | 0.20(1) |
| Acc. | $\rho$ | 0.456(4) | 0.883(3) | 0.706(5) | 0.76(1) | 0.532(8) | 0.489(7) |

The fact that different methods works better for different data set and that the important disease spreaders are harder to predict in some data than others, of course, comes from differences in the temporal network structure. In Table 1, we list values of some structural measures. We see many similarities between the data sets, perhaps because they are all social networks (in the sense that the vertices represent individuals). All data sets have skewed and broad degree distributions (not shown) and they all have bursty contact patterns between along the edges. We will take these observations as guidance when we test our network representations on synthetic data below.

Next, we turn to studying the network structure of the optimized networks of the three types of network representations. The results are shown in Table 4. We include numbers for the accumulated networks for comparison. First, a little side remark—we note that these accumulated networks differ much in structure. The *Gallery* data has much higher assortativity and clustering coefficient than the others. These networks also have longer distances (which is natural because the visitor at the gallery are connected to visitors around the time, so it becomes stretched out in time). The *Conference* data, we note, has a peculiarly high average degree. As for the assortativity (the tendency that high-degree vertices are primarily connected to other high-degree vertices, and low-degree vertices to low-degree vertices), we note that for all types of representations and all empirical networks, except *Gallery*, show negative values. Indeed, the (non-*Gallery*) empirical networks have even more negative values than the null-model (networks with the same set of degrees as the original but otherwise maximally random). This network structure is known to increase epidemic thresholds and the maximal outbreak sizes [27]. The clustering coefficients (normalized number of triangles) are low in absolute numbers but often larger than the null model. Thus, there seems to be a mechanism of the network evolution that promotes the formation of triangles. A high clustering coefficient is known to lower the spreading speed [28]. The average sizes of the largest connected component are typically larger in the empirical network than in the null models (again *Gallery* being an exception). The average distances in the networks are sometimes smaller and sometimes larger than the null-model networks. For networks embedded in space, like the *Prostitution* data, one can expect the empirical network to have larger $d$-values than in the null model. This is indeed the case as can be seen in Tables 4–7. Compared to the network of accumulated contacts, the time-slice and exponential-threshold networks have fewer (non-zero degree) vertices and edges. However, the difference is never larger than 50%. Furthermore, these networks have a size of the largest connected component being close to unity. This is good if we would like to use the derived networks for other types of network epidemiological studies. If the networks would be disconnected, an epidemic simulation could miss possible system-wide outbreaks. Many of these conclusions do not hold for the concurrency networks. These networks are often much smaller— e.g. in the *Dating* data, the numbers of vertices and edges are 761 and 548 respectively, in contrast to the network of accumulated contacts with 22,287 vertices and 78,608 edges. (At the same time, it is not the case that more edges necessarily are better, as we know from the performance of the accumulated networks.) This sheds a new light on the poor performance of the concurrency network representation in the empirical data sets—there is a too low fraction of agents with concurrent partnerships for these to be efficient. However, for the *E-mail 2* data, the performance is actually even better in the concurrency than the time-slice networks. This is also the data set with the largest fraction of concurrent relationships. Therefore, in our case, even if concurrent partnerships increases the importance in disease spreading, they are less significant than accumulated serial contacts (as captured by the time-slice networks). In sum, both the time-slice and exponential-threshold models do not change the structure of the network (compared to the networks of accumulated edges) in any systematic way, but the concurrency networks do.

**Table 3.** Maximal performance for the empirical data sets using coreness as importance measure for the static networks. The format is the same as Table 2.

| | | E-mail 1 | E-mail 2 | Dating | Gallery | Conference | Prostitution |
|---|---|---|---|---|---|---|---|
| Time slice | $\rho_{max}$ | 0.739(1) | 0.907(2) | 0.829 | 0.77(2) | 0.778(3) | 0.731(2) |
| | $t_{start}$ | 0.0 | 0.0 | 0.0 | 0.0 | 0.0 | 0.0 |
| | $t_{stop}$ | 0.42(2) | 0.25(3) | 0.65(3) | 0.72(3) | 0.10(2) | 0.77(3) |
| Concurrency | $\rho_{max}$ | 0.496(4) | 0.912(5) | 0.419(3) | 0.53(2) | 0.759(3) | 0.301(3) |
| | $t_{start}$ | 0.27(2) | 0.17(3) | 0.25(2) | 0.39(3) | 0.07(2) | 0.60(3) |
| | $t_{stop}$ | 0.27(2) | 0.17(2) | 0.75(3) | 0.39(3) | 0.10(1) | 0.60(3) |
| Exponential threshold | $\rho_{max}$ | **0.775(4)** | **0.930(2)** | **0.868(2)** | **0.87(2)** | **0.780(2)** | **0.742(2)** |
| | $\tau$ | 0.40(2) | 1.0(1) | 0.20(1) | 0.64(5) | 0.60(5) | 0.032(4) |
| | $\Omega$ | 0.30(2) | 0.26(1) | 0.20(2) | 0.78(3) | 0.22(2) | 0.22(2) |
| Acc. | $\rho$ | 0.459(4) | 0.884(3) | 0.721(4) | 0.76(1) | 0.394(6) | 0.522(5) |

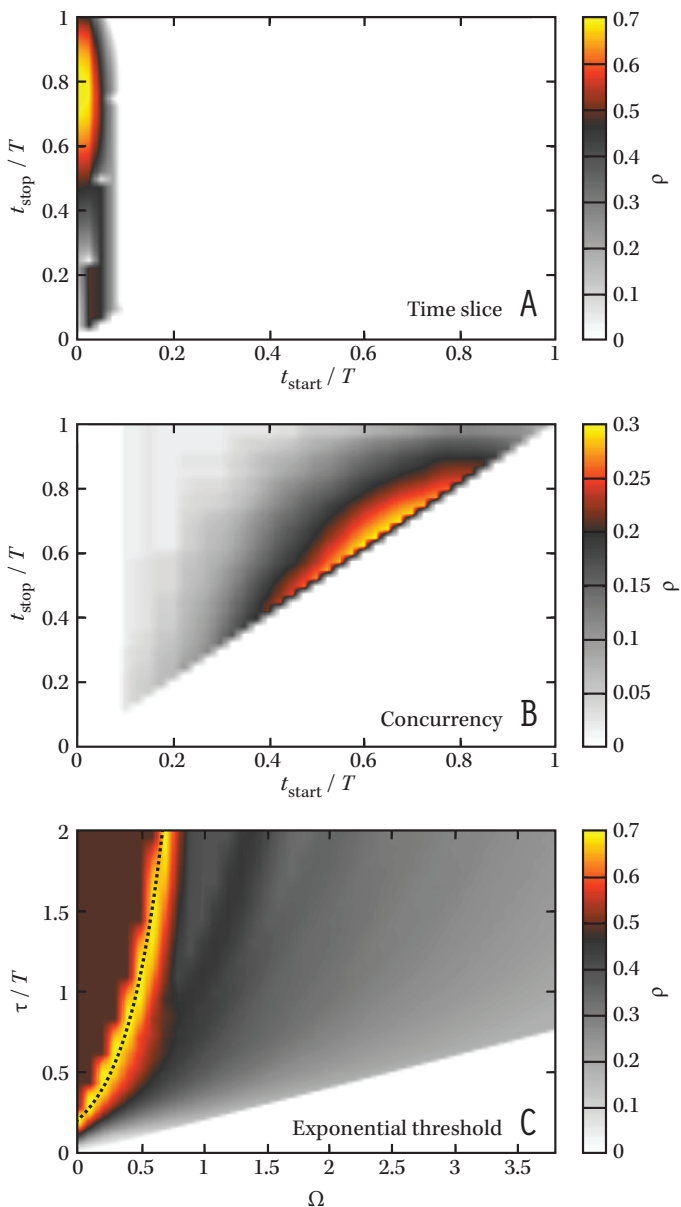

**Figure 2.** The performance of the network representation ρ as a function of parameter values for the *Prostitution* data and degree as an estimate of static-network importance. Panel A shows data for the time-slice networks; B displays results for the concurrency networks and C gives the picture for the exponential-threshold representation. The dotted line illustrates the exponential form of the region of optimality (the equation being $\tau / T = 2e^{\Omega / 0.32}$). The quantities of dimension time are, as indicated, rescaled by the total sampling time $T$ (2,232 days in this case).

### Synthetic networks

Now we will explore effects of the temporal-network structure and the stability of the above observations in a model network. It would be quite impossible to scan all facets of temporal-network structure. Rather, we will focus on the effect of overlapping relationships on the performance of the representations. Can it be the case that they are outperforming the time-slice and exponential-threshold networks for some temporal-networks with a high degree of overlapping relationships? We set up the simulation so as to mimic as much of the observed structure as possible, while simultaneously controlling the average fraction of concurrent relationships. The latter is achieved through a parameter, $\mu \in (0,1]$, where larger values mean more relationships that are concurrent. An outline of the construction algorithm is shown in Fig. 3; for more details about the simulation, see the Methods section.

In Fig. 4, we plot the performance (same as before—the maximum of the Spearman rank correlations between $\Sigma_i$ in SIR simulations and the degree, or coreness, of the respective static network) as a function of μ. As expected, the concurrency networks works better for larger values of μ, but they are never able to catch up with the time-slice and exponential-threshold networks. The difference between the latter two representations is—just like for the empirical networks—small, but with an edge to the exponential-threshold networks. For the largest value of μ, the time-slice networks perform slightly better (but the values are within one standard deviation from one another). Just as for the empirical networks, the degree and coreness are roughly equally good in measuring importance. More precisely, coreness gives higher $\rho_{max}$ for all network representations and measured μ-values, but never more than 4%.

When $\mu = 1$, in the limit of many contacts per edge, the concurrency and time-slice networks will be the same (simply equaling the network of aggregated contacts). The difference, seen in Fig. 3, is because we have on average just 10 contacts per edge. To explore the difference in topology a bit further, we plot the number of vertices of degree larger than zero and average degree in Fig. 4A and B. As expected, when μ is large, these two quantities are quite similar for all network representations. For a lower fraction of concurrent partnerships, however, both the size and the average degree are considerably smaller for the concurrency networks. Like the empirical networks, it seems that the concurrency network representation is too restrictive in its edge definition. Another phenomenon observed in the empirical data that is also reproduced by the synthetic data, is that the networks have larger sizes of the largest connected components than be expected from a randomized null-model (see Fig. 4C and D). This means that the optimized networks have a bias for being more connected. Probably, this reflects that the performance measure relates the local network structure to the outbreak size. I.e., by constructing network whose local properties (degree and coreness) encode a global dynamic property (outbreak size) of the original data, one also affects a global topological property (size of the largest connected component) of the constructed network.

# DISCUSSION

We have explored how to encode as much information from a temporal network and a known start time of an infection into static graphs so that two predictors of disease-spreading importance—degree and coreness—are as accurate as possible. The main conclusions are that, on one hand, exponential-threshold networks generally perform best; on the other hand, time-slice networks often perform almost as good. Our general recommendation is thus to use exponential-threshold networks if possible. However, the simplicity in constructing and optimizing a time-slice network makes it a feasible alternative. To straightforwardly use a network of accumulated contacts is not a good idea—for some data sets, the performance is less than 60% of the maximum. In addition, the concurrency networks—recording contacts that are active simultaneously—perform rather poor. The performance is better when there are relatively many concurrent partnerships (i.e. when these networks are rather dense), but never as good as the other two methods. It is well established that the overall level of concurrent partnerships increases the frequency of population-wide outbreaks [19,20], but it seems like, at least in our data sets, the non-concurrent contacts are necessary for determining the importance of individuals in the spreading process.

How much do our results generalize beyond our current analysis? There are of course many other ways to evaluate the performance of network representations. Instead of the performance measure that we consider (the ability of a vertex' degree, or coreness, to predict its rank in a list of estimated sizes of outbreaks originating at that particular vertex), one can imagine other measures. Different types of centrality measures [29] are candidates for such measures, but these are often global quantities. In practical applications, it is hard to assess quantities other than local—cf. it is easy to check one's degree in an online social network as Facebook, but much harder to know one's coreness (and then coreness is best described as in between a local and global property). Moreover, in many empirical networks, centrality measures (including degree) are strongly correlated in empirical data [30], so we ex-pect our conclusions to remain if we change the static-network importance estimators. Furthermore, the importance measure for the dynamic simulation can be chosen differently. It measures expected outbreak size if the outbreak *starts* at the focal vertex. In general, another factor affecting the importance is the chance to acquire the infection. Ideally, an importance measure should weigh together both these aspects. In most types of data, these aspects are be strongly correlated and we settle for the mentioned expected outbreak size. One can also think of other prediction tasks for the comparison than finding influential spreaders—for example, predicting epidemic threshold, peaktime of the epidemics, prevalence as a function of time or the final outbreak size. Such studies would require us to formulate a disease-spreading model for the static network. This added complication is the main reason that we avoid such a direction. However, we also believe (as mentioned above), that predicting influential spreaders is a comparatively easy task. If one cannot say who would be an influential spreader, but still get the epidemic threshold right, the latter seems rather like luck. (Investigating this hypothesis rigorously would be an interesting future direction.) How much do our conclusions depend on the disease simulation model and its parameter values? The per-contact transmission probability probably does not affect the ranking of the vertices (even if the expected outbreak sizes can vary non-linearly). The duration of the infective state, however, could change the ranking. If the duration is longer, then we anticipate contacts over a longer time span to matter. The network representations should of course be adapted to such a change, in the sense that their optimal parameter values would change. It is hard to see why this would change the ranking of the representations, and a preliminary study (investigating the *Prostitution* data for other δ-values) shows it does not. Other studies [10,31] also find that qualitative results, like the ranking of influential individuals, are robust to the choice of compartmental model and parameter values. The maybe most serious reason to be cautious about generalizing our results is that we have investigated only a limited set of temporal-network structures. Indeed one can imagine numerous types of correlations between temporal structure and network position—correlations between edges connected to the same vertex, between vertices connected by an edge, etc. A promising sign, however, is that the empirical data sets span a rather large range of static network structure (both in terms of the network of accumulated contacts and the optimized networks). In the end, it is probably impossible to scan all temporal-network structures. Rather, we hope for higher quality empirical data. This would also allow us to better tailor the network repre-sentations to specific pathogens.

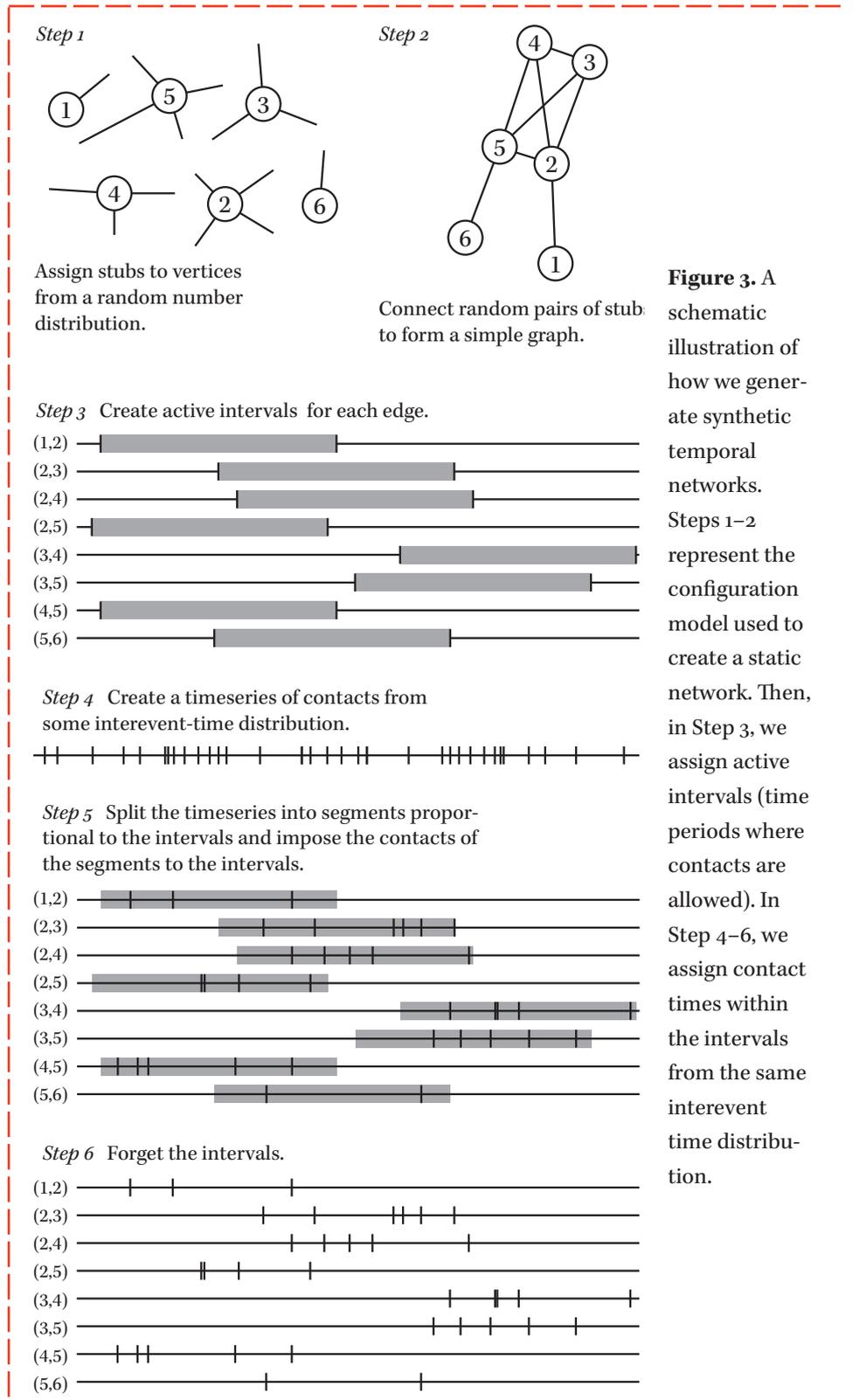

**Figure 3.** A schematic illustration of how we generate synthetic temporal networks. Steps 1–2 represent the configuration model used to create a static network. Then, in Step 3, we assign active intervals (time periods where contacts are allowed). In Step 4–6, we assign contact times within the intervals from the same interevent time distribution.

**Table 4.** Network properties of the optimized networks (with respect to degree as importance predictor) of the empirical contact sequences. $S$ is the fraction of vertices in the largest connected component. $d$ is the average pathlength in the largest connected component. $r$ is Newman's assortativity and $C$ is the clustering coefficient. The italicized numbers are values from a reference model with the same degree sequence as the original network, but otherwise being random. These are averages over $10^4$ randomizations and all the digits are significant to one standard deviation. The values for the *Gallery* data are averaged over the 69 days of sampling. The standard deviations of these mean values are indicated in the parentheses in the same way as in Tables 1, 2 and 3.

| | | E-mail 1 | E-mail 2 | Dating | Gallery | Conference | Prostitution |
|---|---|---|---|---|---|---|---|
| Time slice | N | 25,995 | 2,752 | 23,941 | 132(7) | 84 | 10,958 |
| | M | 38,938 | 18,324 | 93,348 | 545(49) | 531 | 22,095 |
| | S | 0.982 | 0.997 | 0.975 | 0.87(2) | 1 | 0.934 |
| | | *0.676* | *0.934* | *0.847* | *0.984(2)* | *1.00* | *0.750* |
| | d | 3.70 | 2.86 | 4.07 | 3.7(1) | 1.94 | 5.95 |
| | | *3.63* | *2.83* | *3.71* | *2.69(4)* | *2.05* | *4.36* |
| | r | −0.163 | −0.228 | −0.053 | 0.31(3) | −0.117 | −0.117 |
| | | *−0.068* | *−0.055* | *−0.010* | *−0.034(3)* | *−0.065* | *−0.012* |
| | C | 0.00019 | 0.044 | 0.0063 | 0.46(2) | 0.305 | 0 |
| | | *0.0052* | *0.027* | *0.0042* | *0.088(6)* | *0.125* | *0.0040* |
| Concurrent | N | 9,787 | 2,245 | 761 | 28(1) | 78 | 867 |
| | M | 12,494 | 10,558 | 548 | 51(5) | 286 | 784 |
| | S | 0.847 | 1 | 0.293 | 0.51(3) | 1 | 1 |
| | | *0.579* | *0.924* | *0.432* | *0.91(1)* | *0.99* | *0.99* |
| | d | 5.25 | 2.99 | 4.98 | 1.9(1) | 2.42 | 7.32 |
| | | *4.10* | *2.94* | *2.77* | *2.7(1)* | *2.43* | *4.84* |
| | r | −0.090 | −0.224 | −0.143 | 0.44(4) | −0.113 | −0.179 |
| | | *−0.014* | *−0.047* | *−0.067* | *−0.12(1)* | *−0.036* | *−0.023* |
| | C | 0.0090 | 0.037 | 0.0016 | 0.66(2) | 0.288 | 0 |
| | | *−0.015* | *0.023* | *0.0047* | *0.20(2)* | *0.110* | *0.0058* |
| Exponential threshold | N | 31,451 | 2,357 | 22,287 | 147(7) | 110 | 10,566 |
| | M | 47,949 | 12,856 | 78,608 | 455(43) | 864 | 20,390 |
| | S | 0.984 | 1 | 0.963 | 0.57(4) | 1 | 0.924 |
| | | *0.654* | *0.932* | *0.833* | *0.94(1)* | *1.00* | *0.736* |
| | d | 3.77 | 2.93 | 4.29 | 3.7(2) | 2.01 | 6.00 |
| | | *3.69* | *2.86* | *3.81* | *3.21(8)* | *2.04* | *4.38* |
| | r | −0.132 | −0.258 | −0.062 | 0.48(3) | −0.159 | −0.118 |
| | | *−0.055* | *−0.063* | *−0.010* | *−0.031(2)* | *−0.048* | *−0.012* |
| | C | 0.0003 | 0.035 | 0.0075 | 0.54(2) | 0.337 | 0 |
| | | *0.0042* | *0.027* | *0.0045* | *0.062(3)* | *0.135* | *0.004* |
| Accumulated | N | 57,189 | 3,188 | 28,972 | 159(8) | 113 | 16,730 |
| | M | 92,442 | 31,857 | 115,684 | 647(57) | 2,196 | 39,044 |
| | S | 0.999 | 1 | 0.977 | 0.81(3) | 1 | 0.945 |
| | | *0.989* | *0.997* | *0.880* | *0.980(2)* | *0.990* | *0.792* |
| | d | 3.93 | 2.78 | 4.05 | 4.17(13) | 1.66 | 5.78 |
| | | *4.10* | *2.85* | *3.79* | *2.83(3)* | *1.70* | *4.36* |
| | r | −0.081 | −0.258 | −0.048 | 0.37(3) | −0.123 | −0.110 |
| | | *−0.020* | *−0.041* | *−0.0036* | *−0.027(2)* | *−0.049* | *−0.0029* |
| | C | 0.001 | 0.058 | 0.0060 | 0.47(2) | 0.495 | 0 |
| | | *0.001* | *0.016* | *0.0014* | *0.067(3)* | *0.279* | *0.002* |

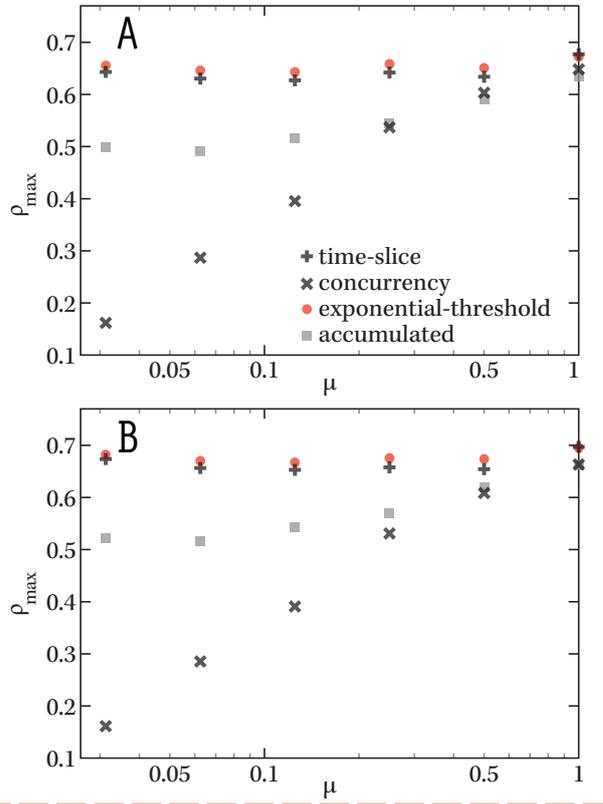

**Figure 4.** Performance of the network representations on the synthetic data sets. We display the maximum value of the Spearman rank correlation as a function of the overlap parameter μ (a model-parameter controlling the fraction of concurrent relationships). Panel A shows the values for degree as importance measure; Panel B is the corresponding for coreness as importance measure. Error bars showing the standard error would be smaller than the symbol size and are not plotted.

The $\rho_{max}$-values—between 0.68 (for the synthetic data) to 0.93 (for the *E-mail 2* data)—are, on one hand, rather high (we could predict important spreaders with a quite high confidence). On the other hand, since many goals of network epidemiology (predicting thresholds, etc., as listed above) are more dependent on the details of the contact structure and thus more difficult, we can appreciate the value of having the full, temporal contact patterns. The conclusion from this is to, as long as possible, avoid reducing contact data to static networks [10,31–36].

An interesting question for the future is why some data sets give higher performance values. With the degree sequences of the accumulated networks $\rho_{max}$ is bounded above by about 0.95–0.98 (1 is unattainable because of the degeneracy of degrees). The discrepancy comes from the network-construction methods being too blunt to capture the relevant temporal-network structure. On the other side, it may be too much to ask from the method to rank the bulk of peripheral vertices accurately—the difference between them will probably be smaller than the errors in the raw data set. Another open future direction is to design other network representations, perhaps putting different weight depending on burstiness [37], "dynamic strength" [38] or other temporal traits.

## METHODS

### Notations

We consider a set $C$ of $L$ contacts among $N$ vertices. $T$ is the total sampling time. We count time (usually denoted $t$) from the data set's first contact. $E$ is the set of vertex pairs with at least one contact. In the context of concurrency, we also call edges "partnerships" to conform to the terminology of the theory of sexually transmitted infections. By construction of our data sets, all the vertices will be part of at least one contact. We denote the number of elements in $E$ by $M$. When we discuss the constructed networks, we use $N$ and $M$ to represent the number of vertices and edges, respectively, in that particular network.

### Disease spreading dynamics

We simulate disease spreading by a version of the SIR model defined as follows. Start the simulation from a situation where all vertices are susceptible. The outbreak is then initiated from a seed $i$ at the time of $i$'s first contact. Then, at every contact involving one infective and one susceptible, we make the susceptible infective with a probability $\lambda$. An individual stays infective for a duration $\delta$, whereupon it becomes removed. (This is different from the differential equation formulations of the SIR model that assumes that infective vertices become removed at a fixed rate—i.e. with an exponentially distributed duration—which is neither realistic [39] nor parsimonious in an individual-based simulation like ours.) We go through the contacts in time order. If more than one contact occurs at a time unit, we sample them in random order. For every vertex as seed, we run the simulation between 1,000 and 10,000 times.

Ideally, we should scan the entire $(\lambda,\delta)$ parameter space, but this would be computationally too demanding. Rather, we will try to simulate the disease spreading where it is easy to separate the more from the less important individuals. This happens at intermediate λ- and δ-values. (For an infinite system, it would be around the epidemic threshold, but for the finite systems that we consider, thresholds are ill defined, so we avoid that terminology.) As a simple principle, we chose δ as one fifth of the sampling time and λ such that the average outbreak size becomes one fifth of the size with $\lambda = 1$ and $\delta = T / 5$. The actual values that we use can be found in Table 1.

### Network representations

We limit ourselves to simple graphs (unweighted and undirected graphs that have no multiple edges or self-edges) and require that their construction should be conceptually simple. The simplest type of such representations is the *time-slice network*—an edge in these is any pair of vertices $(i,j)$ that have one or more contacts $(i,j,t)$ with $t_{start} \leq t \leq t_{stop}$ [16,17]. If $t_{start}$ and $t_{stop}$ are the beginning and end of the data set, then we speak of an aggregated network (which probably is the most common representation when running disease simulations on empirical network data [3]). The second network representation that we consider is *concurrency networks*. Here an edge represents a pair of vertices $(i,j)$ that have at least two contacts $(i,j,t)$ and $(i,j,t')$ where $t < t_{start} \leq t_{stop} < t'$. Ref. [18] studied these in the special case $t_{start} = t_{stop}$.

The last type of network representation that we test is *exponential-threshold* networks. In these, each pair of vertices is assigned a weight

$$\omega_{ij} = \sum_{(i,j,t)\in C} \exp(-t/\tau) \qquad (1)$$

and if this weight exceeds a threshold $\Omega$, then $(i,j)$ is considered an edge. The motivation for this type of network is that contacts that are further from the introduction of the disease (which in our case happens early in the sampling period) should be less important. The time-slice networks can also include edges by a decaying func-tion of time, only that the function is discontinuous. The exponential weight is a smoother way to account for this decrease of importance (so that many later contacts can equal a few recent contacts).

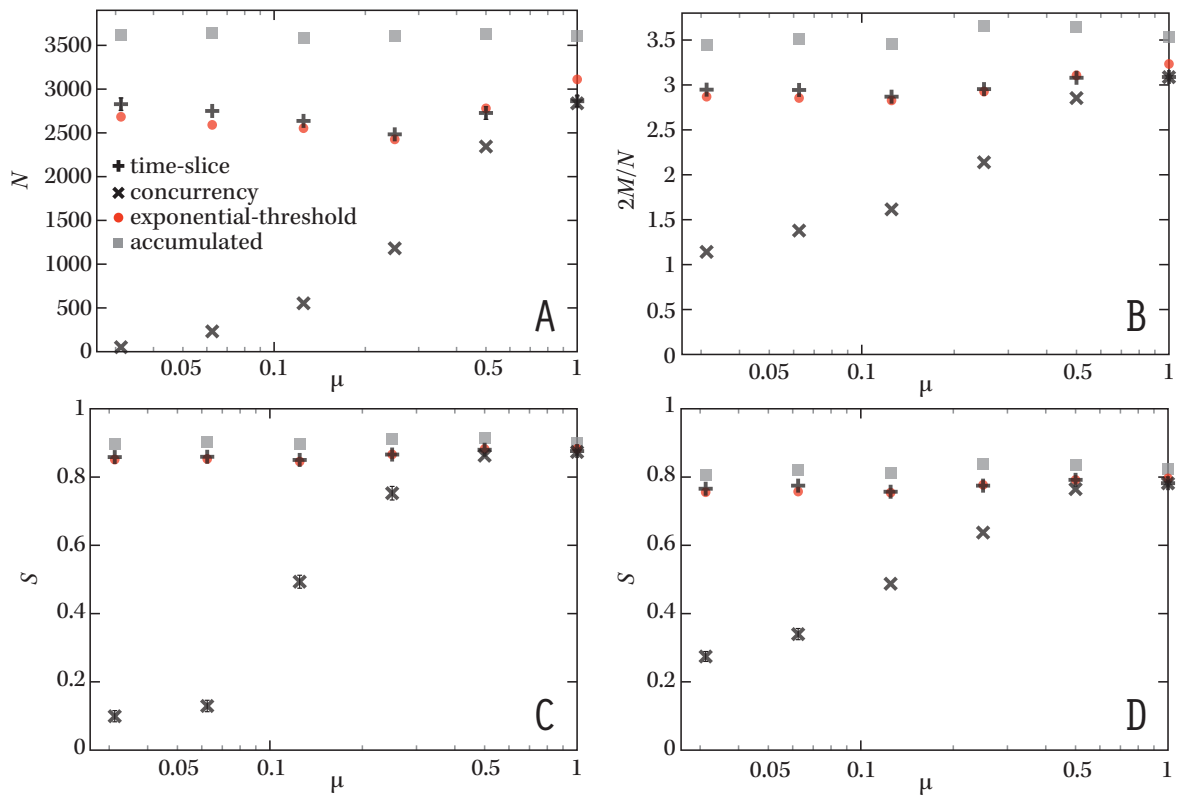

**Figure 5.** Topology of the optimized network representations for synthetic data and degree as a proxy for static-networks importance. Panel A shows the number of (non-zero degree) vertices in the network; B displays the average degree; C gives the relative size of the largest connected component; while D shows the corresponding figure to C for null models of the same degree sequences as in C, but otherwise random. The error bars are displayed if they are about the same size as the symbols and show the standard error.

**Quantities to characterize the temporal-network structure**

In our tables discussing the structure of the data sets and derived networks, we use a number of quantities that we will define here.

To quantify the tendency of contacts to be temporally separated by broadly distributed intervals, we use the burstiness measure of Ref. [40]. For all pairs of vertices $i$ and $j$ with two or more contacts, we collect the times between contacts to one long series of interevent times. Then the *burstiness B* of the data set is the coefficient of variation of this series.

To measure the tendency of high-degree vertices to connect to other large-degree vertices and low-degree vertices to connect to low-degree vertices, i.e. the assortative mixing by degree, we use Newman's *assortativity* [27,29]. This is essentially Pearson's correlation coefficient of the degrees of vertices at either side of an edge. The only difference is that since an edge is symmetric, but a list of pairs of degrees $\{(k_i,k_j)\}$ corresponding to the edges is not; we let each edge $(i,j)$ contributes twice to the list—with both $(k_i,k_j)$ and $(k_j,k_i)$.

We measure the tendency of connected triples of vertices to also form a triangle by the *clustering coefficient C* [29]. It is defined as the ratio of the number of triangles and the number of connected triples times a factor 3 (to normalize the quantity to the interval [0,1]).

Another important quantity is the *relative size of the largest connected component* (i.e. a subgraph where every vertex is reachable by following a sequence of adjacent edges). We measure it as a fraction $S$ of the total number of vertices in the graph. Note that this, when applied to the optimized networks, is not necessarily the same as the total number of individuals in the original data set. Finally, we measure the distances in the largest connected component, $d$—the number of edges in the shortest path between two vertices in the largest connected component, averaged over all its pairs of vertices.

We compare the static network measures by the corresponding values from a randomized null model with the same set of degrees but otherwise no structure. An instance of this model is generated by: sequentially going through all edges $(i,j)$, pick a random new edge $(i',j')$, replace these two edges by $(i,j')$ and $(i',j)$, or (with the same probability) by $(i,i')$ and $(j,j')$. If the replacement step would introduce a multiple or self-edge another edge, then a new $(i',j')$ is chosen randomly.

**Predictors of importance**

We estimate importance of a vertex $i$ in the SIR simulation as the average outbreak size $\Sigma_i$ over 1,000 to 10,000 independent simulation runs if the disease is introduced by the vertex at its first contact.

To estimate the importance of a vertex in the disease spreading from the static networks, we consider two quantities that both have been used in the literature for this purpose. The first quantity is *degree* $k_i$—the number of neighbors of a vertex. This is a useful quantity for its simplicity. It is local, meaning that every individual should be able to estimate its own value (in practice this could of course be difficult, depending of the mode of transmission of the pathogen). Degree, as a measure of influence, is also intuitive—meeting more individuals should increase both the chance of getting a disease and the number of others one can spread the infection to.

The second measure we try is *coreness* $c_i$. This is defined through a technique to categorize the vertices into "shells" or "cores" (the terminology is ambiguous). Start by removing all vertices of degree zero (there will be none, by construction, in our case). Then remove all vertices with degree one. If any vertex gets degree one through its neighbor being removed, then remove this vertex too. Continue until there are no vertices of degree one. Then do the same with vertices of degree two, three, etc. The coreness of $i$ is the stage when it is removed (one, if it is removed when starting to remove vertices of degree one, and so on). Ref. [8] argues (supported by simulations) that this is a better measure of importance than degree since a vertex with high coreness is member of a cluster of vertices that can sustain a pathogen better than e.g. the surrounding of a lonely high-degree vertex.

**Generative models for contact sequences**

The method to generate synthetic contact sequences is outlined in Fig. 3. Here we describe the process in greater detail. We start by constructing a (static) simple graph, $N = 5{,}000$, by the configuration model [27]. This means that we assign one discrete random number for every vertex $i$ from some probability distribution. These numbers represent "stubs" or "half-edges" desired to be a part of an edge. Then we choose stubs of random pairs of vertices $i$ and $j$ and attach them to form an edge $(i,j)$ provided that no such edge already existed and $i \neq j$. This adding of edges continues until there is no stub that is not a part of an edge. One caveat, however, is that a complete matching may not be possible (if, for example, one vertex has two stubs left and the others all zero). To handle this, if the matching is unsuccessful for $10^4$ consecutive times, we give up and delete the remaining stubs. In this paper, we use a truncated power-law distribution to mimic the skewed, broad degree distributions of the empirical networks. To be specific, we draw the random numbers from a distribution

$$P(k) = \begin{cases} k^{-\gamma} & \text{if } k_{\min} \leq k \leq k_{\max} \\ 0 & \text{otherwise} \end{cases} \quad (2)$$

where, in our work, $k_{\min} = 1$, $k_{\max} = N - 1$ and $\gamma = 2.2$. This gives, on average, $M = 10{,}595(5)$ (the last number being the standard error in order of the last decimal).

After the network topology is generated, we proceed to assign times of contacts to the edges. We assume a contact over an edge can only take place during an activity interval of duration, $\mu T$. We recognize that the activity intervals would more accurately be modeled as skewedly distributed. However, if we choose the intervals as broadly distributed as e.g. the degrees, then we cannot control the overlap of contacts over such wide a range. $\mu \in (0,1]$ is then a control parameter for the overlap with larger values meaning a higher amount of overlap. (Note that $\mu$ increases with the average fraction concurrent relationships, but to avoid confusion by the concurrency measure of Ref. [20], we do not call $\mu$ concurrency.) The starting times of the intervals are chosen with uniform probability in the interval $[0,(1 - \mu)T]$.

We proceed by generate a time series with, once again, a truncated power-law shape. We use the equation

$$P(\Delta) = \begin{cases} \Delta^{-\beta} & \text{if } \Delta_{\min} \leq \Delta \leq \Delta_{\max} \\ 0 & \text{otherwise} \end{cases} \quad (3)$$

where $\Delta_{\min} = 1$, $\Delta_{\max} = 10^4$ and $\beta = 2$. We generate $L = 10M$ such contacts. This times series is then split over the active intervals. When that is finished, the temporal network is done. Note that this procedure does not induce any particular correlations between topology and temporal structure. Ref. [41] uses a similar method that differs in that it does not assign active intervals (and thus does not have the control parameter $\mu$).

**Acknowledgments**

We are grateful for helpful comments from Fariba Karimi, Naoki Masuda, Luis E. C. Rocha and Jari Saramäki. This work was supported by the World Class University Program from NRF, Korea (R31–2008–10029), and the Swedish Research Council.